\documentclass[10pt,a4paper]{article}

\usepackage{graphicx}
\graphicspath{{Plots/}}
\usepackage{url}
\usepackage{latexsym}
\usepackage{a4wide}

\newcommand{\inet}{{\em \sc inet}}
\newcommand{\web}{{\em \sc web}}
\newcommand{\ptp}{{\em \sc p2p}}
\newcommand{\ip}{{\em \sc ip}}

\newcommand{\ecc}{\ensuremath{{\mbox{ecc}}}}
\newcommand{\diam}{\ensuremath{D}}

\newcommand{\dslbs}{d.s.l.b.}
\newcommand{\hdtubs}{h.d.t.u.b.}
\newcommand{\tlbs}{t.l.b.}
\newcommand{\tubs}{t.u.b.}
\newcommand{\rtubs}{r.t.u.b.}
\newcommand{\dslb}{double sweep lower bound}
\newcommand{\Dslb}{Double sweep lower bound}
\newcommand{\hdtub}{highest degree tree upper bound}
\newcommand{\tlb}{trivial lower bound}
\newcommand{\tl}{trivial lower}

\newcommand{\tub}{trivial upper bound}
\newcommand{\rtub}{random tree upper bound}
\newcommand{\treeub}{tree upper bound}
\newcommand{\Treeub}{Tree upper bound}

\newcommand{\degree}[1]{\ensuremath{d^\circ(#1)}}

\newcommand{\qed}{\hfill$\Box$ \vspace{0.5 cm}}

\newcommand{\ie}{{\em i.e.}}
\newcommand{\polylog}{\ensuremath{\mbox{polylog}}}

\newcommand{\plots}[3][0.6]{\begin{figure}[!h]
\includegraphics[scale=#1]{Plots/Inet/#2.fig.eps}
\hfill
\includegraphics[scale=#1]{Plots/P2P/#2.fig.eps}\\
\includegraphics[scale=#1]{Plots/Web/#2.fig.eps}
\hfill
\includegraphics[scale=#1]{Plots/IP/#2.fig.eps}
\caption{#3
From left to right and top to bottom: \inet, \ptp, \web\ 
and \ip\ graphs.}
\label{fig_#2}
\end{figure}}

\begin{document}

\begin{center}
{\Large \bf
Fast Computation of Empirically Tight Bounds\\
\smallskip
for the Diameter of Massive Graphs
}\\
\medskip
Cl\'emence Magnien\,\footnote{%
LIP6 -- CNRS and UPMC ({\tt firstname.lastname@lip6.fr})},
Matthieu Latapy\,\footnotemark[1] and
Michel Habib\,\footnote{%
LIAFA -- CNRS and Universit\'e Paris Diderot ({\tt firstname.lastname@liafa.jussieu.fr})}
\end{center}

\footnotetext[3]{
\copyright{} ACM, (2009). This is the author's version of the work. It is posted here by permission 
for your personal use. Not for redistribution. The definitive version was published in
JEA, \{19, 1084-6654, (February 2009)\} \\
\url{http://portal.acm.org/citation.cfm?doid=1412228.1455266}
}

\bigskip
\begin{abstract}
The diameter of a graph is among its most basic parameters.
Since a few years, it moreover became a key issue to compute it for massive graphs in the context of complex network analysis. However, known algorithms, including the ones producing approximate values,
have  too high a time and/or space complexity to be used in such cases.
We propose here a new approach relying on very simple and fast algorithms that compute (upper and lower) bounds for the diameter.
We show empirically that, on various real-world cases representative of complex networks studied in the literature,
the obtained bounds are very tight (and even equal in some cases). This leads to rigorous and very accurate estimations of the actual diameter in cases which were previously untractable in practice.
\end{abstract}

\section{Context.}
\label{sec-art}
\label{sec-context}

Throughout the paper, we consider a connected undirected unweighted graph $G = (V,E)$ with $n = |V|$ vertices and $m=|E|$ edges. We denote by $d(u,v)$ the distance between $u$ and $v$ in $G$, by $\ecc(v) = \max_u d(v,u)$ the eccentricity of $v$ in $G$, and by $\diam = \max_{u,v} d(u,v) = max_v \ecc(v)
$ the diameter of $G$.

Computing all distances from one vertex to all the others (and thus its eccentricity) has $\Theta(m)$ time and space costs using a breadth-first search (BFS). In order to compute the diameter, one has to compute the distance between all pairs of vertices, which therefore has a $\Theta(n\cdot m)$ time and a $\Theta(m)$ space costs using $n$ BFS.
Using matrix products, one may achieve this in $O(n^{2.376}\cdot\polylog(n))$ time and $\Theta(n^2)$ space \cite{alon1992boolean,seidel1992apsp}.
Therefore, the BFS approach is too slow for massive graphs, and the matrix approach remains too slow and has in addition a prohibitive space cost.

Methods have been proposed to compute the distances between all pairs of vertices without resorting to matrix products but faster than with BFS; see \cite{feder1991clique} for such an algorithm,
in $\Theta\left(\frac{n^3}{log n}\right)$ time and ${\cal O}(n^2)$ space for dense graphs.
See also~\cite{chan2006allpairs} for an algorithm in ${\cal O}(n^2 (\log \log n)^2/\log n)\subset { o}(n^2)$  time
and ${\cal O}(n^2)$ space for sparse graphs.
Again, such algorithms are too slow and too space consuming for practical use on massive graphs.

In \cite{aingworth96diameter,dor97apasp}, the authors propose to compute {\em almost shortest paths} between all pairs of vertices, leading to an estimation $\overline{\diam}$ of $\diam$ such that $\overline{\diam} \le \diam \le \overline{\diam} + 2$.
The cost of this computation however is in 
$\Omega(n^2)$ time
 and $\Theta(n^2)$ space, which is still too  expensive in our context.
See \cite{zwick01exact} for a survey on the computation and approximation of diameters (and distances in general).

The authors of \cite{Parnas2002testingdiameter} use a different approach:
they designed an algorithm for testing probabilistically whether the diameter of 
a graph $G$ is below a given value $D$,
or whether $G$ is {\em close} to such a graph,
in terms of the number of edges to add or remove to transform one graph into the other.

Some authors also designed methods for the approximation or exact computation of the diameter
for some specific classes of graphs~\cite{handler1973tree,CDHP01,corneil2003diameter,dragan1997lbfs,dragan2005unified,chepoi1994central}, which we will use in the following.

\medskip

Finally, current algorithms available to compute the diameter of a given graph have too high a computational cost;
it is in practice impossible to obtain the exact value, or accurate estimations,
of the diameter of massive graphs such as the ones encountered in complex network studies (which typically have hundreds of thousands vertices, and up to a few billions). Because of this, authors give ad-hoc estimations of the diameter (typically obtained as the maximal eccentricity found for a given set of randomly chosen vertices), for which no rigorous assessment is done. Some authors also make no distinction between the diameter and the average distance, which they approximate using various heuristics.

A more rigorous approach consists in replacing the notion of diameter by other notions which capture similar properties while being easier to evaluate. In \cite{leskovecdensification} for instance the authors consider the {\em effective diameter},
\ie\ the value such that $90\%$ of the pairs of vertices are at a distance lower than this value. Such approaches are appealing, but they are out of the scope of this paper, in which we show that the actual diameter may be estimated very accurately and very quickly in practice, even in massive graphs.



\section{Computation methods.}
\label{sec-our}

We now present a series of very simple but powerful algorithms that give exact upper or lower bounds for the diameter.
There is however no guarantee on how accurate these bounds are,
except that if they are close from each other then the actual diameter is well estimated.
The interest of this approach relies on the fact that the bounds we obtain are indeed close, which we study empirically in the next section.

\subsection*{Trivial bounds.}
First notice that the eccentricity of any vertex $v$ gives trivial bounds of the diameter:
$\ecc(v) \le \diam \le 2\cdot \ecc(v)$;
we call these {\em \tl{} and  upper bounds}.
They can be computed in $\Theta(m)$ space and time.

\subsection*{\Dslb{}.}
The quality of the trivial bounds obviously depends on the choice of the vertex $v$. In \cite{CDHP01,handler1973tree} it is shown that, on some specific classes of graphs (including trees),
if $v$ is chosen such that $d(u,v) = \ecc(u)$ for a vertex $u$
(\ie\ $v$ is among the vertices which are at maximal distance from $u$), then $\diam = \ecc(v)$.
In these graphs, the diameter may therefore be computed by a BFS from any node $u$ and then a BFS from a node at maximal distance from $u$, thus in $\Theta(m)$ space and time. In the general case,
the value obtained in this way is different from the diameter,
however, it always gives a better lower bound than than the trivial method from the same vertex
(which gives $\ecc(u)$).
We call bounds obtained in this way {\em \dslb s}.


\subsection*{\Treeub{}.}
Now let us notice that the diameter of any spanning (\ie\ containing all the vertices) connected subgraph of $G$ is larger than or equal to the diameter of $G$; therefore, one may obtain upper bounds by considering appropriate subgraphs of $G$.
As we have juste seen, it is easy to compute the diameter of any tree in $\Theta(m)$ time and space \cite{handler1973tree}.
Spanning trees of $G$, in particular the ones having small diameters, therefore are good candidates for this. We will call a {\em \treeub} for the diameter any upper bound obtained as the diameter of a BFS tree from a vertex.
It always is better than the corresponding \tub.



\subsection*{Iterating the bounds.}
Finally, we obtain a set of very simple algorithms (all relying on a few BFS) which give lower and upper bounds of the diameter. Iterating them from different intial vertices makes it possible to obtain tighter and tighter bounds, with a linear cost for each step. In the context of massive graphs, the number of steps must therefore remain small, but it may be worth to make several iterations (as we will see in the next section).

The initial vertices may be chosen randomly, but one may also choose vertices
 more likely to produce tight bounds. We will do so for the \treeub: as highest degree vertices are intuitively likely to produce BFS trees with small diameter, we will consider vertices in decreasing order of degrees when iterating this algorithm.
This is motivated by the fact that real-world complex networks are known to have some vertices with very high degree,
see for instance \cite{watts1998smallworld,faloutsos99sigcomm,broder00graph}.
We call this process the {\em \hdtub{}};
in oppostion, we call iterating the \treeub{} from random vertices the {\em \rtub{}}.


Notice that one may expect that iterating the computations for a long time will eventually lead to equal bounds, and thus to the exact computation of the diameter.
There is no guarantee of this, however, as all the \treeub{}s may be strictly larger than $D$
(if $G$ is a cycle of $n$ vertices, for instance, its diameter is $\lfloor n/2 \rfloor$ and the \treeub\ is $n-1$ whichever vertex one starts from).
On the contrary, the \dslb\ will give the actual value of the diameter in some cases, and thus leads to the exact diameter when it is iterated on all vertices.

\subsection*{Implementation.}
We have implemented the different heursitics described above\,\footnote{The program is available at
\url{http://www-rp.lip6.fr/~magnien/Diameter/}.}.
The program is written in C, so as to be most efficient in central memory usage.
Graphs are represented as adjacency arrays:
to each vertex $v$ is associated its degree \degree{v}
and an array of size \degree{v} containing its neighbours.
A graph with $n$ vertices and $m$ edges is therefore stored in a space equal
to $2n + 2m$: two arrays of size $n$ containing the degrees and the references
to the adjacency arrays, 
and the adjacency arrays, with total size $2m$;
this is close to optimal\,\footnote{Our algorithms require that each link $(u,v)$ is
stored in a space equal to at least 2, $u$ appearing in the adjacency array of $v$
and conversely.
This requires a space of at least $2m$ if one does not resort to compression techniques.}.

The only hardware requirement is that the graph thus stored should
fits in central memory:
if swapping occurs the performances of the program are degraded to such an
extent that it cannot finish in a reasonable amount of time.

\medskip
The main behaviour of the program consists in iterating the
\dslb{} and \hdtub{} until the difference between the best bounds
obtained is lower than or equal to a given threshold value.
The choice of this value may depend on many factors: the graph considered,
the desired quality of the bounds, \ldots{}
For all practical purposes however, setting this value to 5 should
 give a good estimate
of the diameter in a short time.

The program also has an option allowing it to run until the diameter is
estimated with a given precision $p$,
\ie{} until the best upper and lower bounds obtained, $\underline{D}$ and $\overline{D}$,
are such that $(\overline{D} - \underline{D}) / \underline{D} < p$.

Finally, the program gives the possibility to compute any of the five bounds
for a given number of iteration.
These options are useful for comparing the different heuristics.

\medskip
Notice that, though all heuristics have a $\Theta(m)$ time complexity,
they do not have the exact same computation time.
Aproximately, one could say that computing a \dslb{} or \treeub{}
takes twice as much time as computing a \tlb{} or upper bound.
As we will see in the next section, 
which presents experimental results for our heuristics,
this does not play a big role for practical purposes:
the \dslb{} and \hdtub{} indeed provide very quickly a very good
estimate for the diameter.
These are therefore the heuristics to use in practice.

\section{Experiments.}

We tested our heuristics on a wide variety of real-world graphs coming from
different contexts, as well as on different types of random graphs.
In all cases, the \dslb{} and \hdtub{} have yielded in a small number of
iterations a good estimation of the diameter,
and have performed better than the other heuristics.

To illustrate this, as well as compare the behaviours of the different bounds,
we chose to present here detailed experimental results
on a set of four real-world graphs,
which may be considered as representative of the variety of cases met in complex network studies\,\footnote{%
These data sets are described in more details in \cite{latapy2007measurement}.
Because they may be useful for other purposes,
and because they are needed to reproduce our results, we provide them at \url{http://www-rp.lip6.fr/~magnien/Diameter/}.}:\\
{\bf an internet topology graph (\inet)} obtained from traceroutes ran daily in 2005 by {\em Skitter}\,\footnote{\url{http://www.caida.org/tools/measurement/skitter/}}
from several scattered sources to almost one million destinations, leading to $1\,719\,037$ vertices and $11\,095\,298$ edges;\\
{\bf a web graph (\web)} containing the $39\,459\,925$ web pages (vertices) and $783\,027\,125$ links (edges) collected in the {\tt .uk} domain during a measurement conducted in 2005 by {\em WebGraph}\,\footnote{\url{http://webgraph.dsi.unimi.it/}};\\
{\bf a peer-to-peer graph (\ptp)} in which two peers are linked if one of them provided a file to the other in a measurement conducted on a large {\em eDonkey} server for a period of 47 hours in 2004\,\footnote{%
\url{http://www-rp.lip6.fr/~latapy/P2P_data/}},
leading to $5\,792\,297$ vertices and $142\,038\,401$ edges;\\
{\bf a traffic graph (\ip)} obtained from {\em MetroSec}\,\footnote{\url{http://www2.laas.fr/METROSEC/}}
which captured each \ip\ packet header routed by a given router during 24 hours,
two \ip\ addresses being linked if they appear in a packet as sender and destination, leading to $2\,250\,498$ vertices and $19\,394\,216$ edges.

On these graphs, 
we iterated the computation of all bounds:
\tlb{}, \tub{}, \dslb{},  \rtub{} (all from randomly chosen nodes),
and \hdtub{} (from nodes chosen by decreasing order of degrees).
The number of iteration was not the same for all the graphs: it should be as large as possible to provide much insight,
but sufficiently small to run the experiment in a reasonable time.
We finally ran $2\,000$ iterations for \web\ graph, $5\,000$ for \ptp{}, and $10\,000$ for \ip\ and \inet. As we will see, these numbers have little impact on our conclusions, if any.

\begin{table}[!h]
\centering
\begin{tabular}{|l||c|c||c|c||c|c||c|c||c|c|}
\hline
 & \multicolumn{2}{|c||}{\tlbs} & \multicolumn{2}{|c||}{\dslbs}
 & \multicolumn{2}{|c||}{\hdtubs} & \multicolumn{2}{|c||}{\rtubs}
& \multicolumn{2}{|c|}{\tubs}\\
\hline
\hline
\inet{} & \multicolumn{2}{|c||}{29} & \multicolumn{2}{c||}{{\bf 31}}& \multicolumn{2}{|c||}{{\bf 34}} &
\multicolumn{2}{|c||}{34} & \multicolumn{2}{|c|}{38}\\
\cline{2-11}
 &
998 & 0.0001 &
2 & 0.49 &
26 & 0.1002&
23 & 0.0633 &
127 & 0.0011  \\
\hline
\ptp{} & \multicolumn{2}{|c||}{8} & \multicolumn{2}{|c||}{{\bf 9}}& \multicolumn{2}{|c||}{{\bf 10}} &
\multicolumn{2}{|c||}{10} & \multicolumn{2}{|c|}{10}\\
\cline{2-11}
 &
210 &0.0094 &
1 & 0.7005 &
1 & 0.039 &
120 & 0.0032 &
3237 & 0.0001 \\
\hline
\web{} & \multicolumn{2}{|c||}{26} & \multicolumn{2}{|c||}{{\bf 32}}& \multicolumn{2}{|c||}{{\bf 33}} &
\multicolumn{2}{|c||}{33} & \multicolumn{2}{|c|}{34}\\
\cline{2-11}
 &
816 & 0.001 &
1 & 0.985 &
34 & 0.0015 &
46 & 0.0025 &
1572 & 0.0005 \\
\hline
\ip{} & \multicolumn{2}{|c||}{9} & \multicolumn{2}{|c||}{{\bf 9}}& \multicolumn{2}{|c||}{{\bf 9}} &
\multicolumn{2}{|c||}{9} & \multicolumn{2}{|c|}{10}\\
\cline{2-11}
 &
5331 & 0.0001 &
1 & 0.989 &
4 & 0.0543 &
12 & 0.0346 &
6284 & 0.0001 \\
\hline
\end{tabular}
\caption{Best bounds obtained for our four graphs by iterating each computation method: \tlb{} (\tlbs), \dslb{} (\dslbs{}), \tub{} (\tubs{}), \rtub{} (\rtubs{}), and \hdtub{} (\hdtubs{}). In each case, we also give the number of iterations we had to perform to obtain this bound for the first time (left), as well as the frequency with which we obtained this bound (\ie\ the number of times we obtained it divided by the number of iterations we ran) (right).}
\label{tab-results}
\end{table}

%

Table~\ref{tab-results} summarises the results;
it not only gives the results obtained in each case,
but also some additional information on the performance of each method. We will give more details below,
but one may already notice that the obtained bounds are tight even with the simplest computation methods:
the worst case is the one of the trivial lower and upper bounds for \inet,
the best values they provide being respectively $29$ and $38$,
which may already be considered as a relevant information on the actual diameter. The other bounds are tighter, and even equal (to $9$) in the case of \ip, thus giving the exact diameter in this case.

The obtained values are not the only meaningful parameter to evaluate the performance of the various methods; the frequency and the first time at which we obtain the given bound are also precious indicators which we give in Table~\ref{tab-results}.
One may observe for instance that in the case of \web\ not only the best \tlb\ is $26$ and the best \dslb\ is $32$,
which is much better,
but also the best \tlb\ is obtained with frequency $0.001$ while the other is obtained with frequency $0.985$.
In other words, in this case, the best \tlb\ will be obtained approximately once every $1000$ iterations,
while the other, though it is much better, is obtained at almost every iteration.

Similar comments may be made on the \tub\ and the \rtub: in the case of \ip, for instance, one obtains $10$ with frequency $0.0001$ and $9$ with frequency $0.0346$, respectively. The \rtub\ and the \hdtub\ always give the same best value in our experiments,
but the latter gives it with a much higher frequency, and thus with a lower expected number of iterations. Note however that in the case of \web\ the difference is not as clear: though the \hdtub\ reaches the best value rather quickly ($34$ steps), it does so with frequency $0.0015$ while the \rtub\ gives its best value with a frequency of $0.0025$.


\medskip

In order to deepen our understanding of the behavior of each proposed bounding method, we however need to observe not only the best obtained value but all the values obtained during the iteration.
This may be done by studying for each upper bound computation method the {\em cumulative distribution} of the obtained values
(\ie\ for all $k$ the number of times that the obtained lower bound was less than or equal to $k$)
and for each lower bound computation method the {\em complementary} cumulative distribution of the obtained values (\ie\ for all $k$ the number of times that the obtained lower bound was greater than or equal to $k$).

\plots{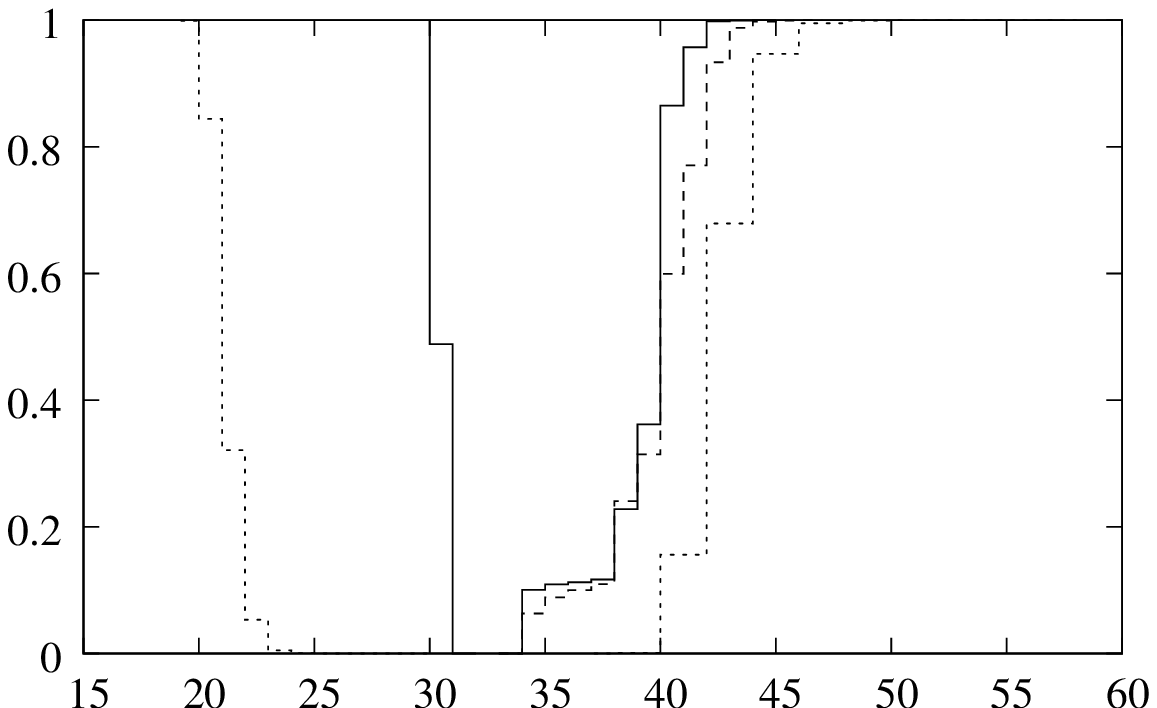}{Cumulative distributions of the bounds obtained with all the different heuristics.
For the lower bounds: \tlb{} (dotted line, left) and \dslb{} (full line, left).
For the upper bounds: \tub{} (dotted line, right), \rtub{} (dashed line, right) and \hdtub{} (full line, right). }

These distributions are presented in Figure~\ref{fig_distr_cum_upper_lower}.
They show for instance that the \tlb\ for \ptp\ gives a bound at least as good as $7$ with probability $0.67$.
Likewise, still for \ptp, the \hdtub\ point at coordinates $(11, 0.53)$ means that it gives a bound at least as good as $11$ with probability $0.53$.

For each heuristic, the best obtained value is given by the point at which the corresponding plot meets the horizontal axis. If a method for the lower bound performs better than another, its plot will be on the right of the other's plot. Conversely, if a method for the upper bound performs better than another, its plot will be on the left of the other's plot.
Notice that the plot for the \dslb{} meets the plots for both \treeub{}s at the value $9$ in the case of the \ip{} graph.
This shows again that the exact value of the diameter was obtained for this graph.

Finally, these plots show that the observations above also hold for the general behavior of the studied methods. Concerning lower bounds, the \dslb\ performs much better than the \tlb. Concerning upper bounds, the same is true for the \rtub\ and the \tub. The \hdtub\ performs in general slightly better than the \rtub.
In the case of \web\ however, we observed that the \rtub\ gives the value $33$ with a higher frequency than the \hdtub. This is not true in general for this graph: the \hdtub\ performs better than the \rtub\ in all other cases (the best value for both these methods is obtained with a probability too small to be observable on this plot). For \ip, the converse is true: though the \hdtub\ gives the best value with a higher frequency than the \rtub{}, for other cases the \rtub{} seems to perform better than the \hdtub{}.



\section*{Conclusion.}

In this note, we describe very simple methods based on BFS to give bounds on the diameter of a graph.
The originality of this approach lies in the fact that there is a guarantee that the actual diameter is within the bounds we find, but there is no guarantee on the tightness of these bounds.
In some cases, they may be very far of each other, and thus of little interest, if any.
We however show empirically, through a representative set of experiments that we discuss, that one may expect the obtained bounds to be very tight in practice, while having a very small computational cost.
In some cases, the obtained bounds may even be equal, thus giving the exact value of the diameter.

Up to our knowledge, this approach is the only known one able to give bounds in a reasonable time and space with such accuracy in many practical cases, including the ones we present.

This work opens several directions for research.
Designing a method for estimating {\em beforehand} the number of iterations to run 
to obtain a good estimation of the diameter would be a very interesting result in
this context.
Our approach is complementary to the results developed in~\cite{Parnas2002testingdiameter},
which provide a way to estimate probabilistically whether a given graph has a diametre below a given value.
Combining these two approaches would probably lead to interesting insight,
and possibly provide a method for estimating where the diameter is situated within the obtained bounds.
In a related manner, another key direction would be to obtain formal results on the tightness of the obtained bounds,
which may be possible for instance in the case of random graphs.
Finally, an interesting improvement to our results would be to remove (resp. add)
edges to the graph under concern so as to obtain a graph from a class for which the exact diameter can be
computed quickly (e.g., chordal graphs);
this might give tighter bounds than the ones we currently obtain.

\medskip
\noindent
{\bf Acknowledgements.}
We thank all the colleagues who provided the data used in this note;
no such work would be possible without their help. We thank
Vincent Limouzy, as well as the anonymous referees, for helpful comments and references.

\bibliographystyle{plain}
\bibliography{biblio}

\end{document}